\documentclass[preprint,12pt]{elsarticle}
\pdfoutput=1

\usepackage{graphicx}
\usepackage{rotating}

\usepackage{amssymb}
\journal{Physics Letters B}

\begin{document}

\begin{frontmatter}



\title{Shell evolution approaching the $N=20$ island of inversion: structure of $^{26}$Na}


\author[Surrey]{G.L. Wilson}
\author[Surrey]{W.N. Catford\corref{cor1}\fnref{corresponding}}
\ead{w.catford@surrey.ac.uk}
\author[LPC]{N.A. Orr}
\author[York]{C.Aa. Diget}
\author[Surrey]{A. Matta}
\author[TRIUMF]{G.~Hackman}
\author[TRIUMF]{S.J.~Williams}
\author[Surrey]{I.C. Celik}
\author[LPC]{N.L. Achouri}
\author[TRIUMF]{H.~Al Falou}
\author[Liverpool]{R.~Ashley}
\author[StMary]{R.A.E.~Austin}
\author[TRIUMF]{G.C.~Ball}
\author[LSU]{J.C. Blackmon}
\author[Liverpool]{A.~J.~Boston}
\author[Liverpool]{H.~C.~Boston}
\author[Surrey]{S.~M.~Brown}
\author[TRIUMF]{D.~S.~Cross}
\author[TRIUMF]{M.~Djongolov}
\author[Toronto]{T.~E.~Drake}
\author[TRIUMF,Mines]{U.~Hager}
\author[York]{S.~P.~Fox}
\author[York]{B.~R.~Fulton}
\author[TRIUMF]{N.~Galinski}
\author[TRIUMF]{A.~B.~Garnsworthy}
\author[Guelph]{D.~Jamieson}
\author[StMary]{R~Kanungo}
\author[Guelph]{K.~G.~Leach\fnref{mines}}
\author[TRIUMF,UWC]{J.~N.~Orce}
\author[TRIUMF]{C.~J.~Pearson}
\author[Mines]{M.~Porter-Peden}
\author[Mines]{F.~Sarazin}
\author[Surrey]{E.~C.~Simpson}
\author[TRIUMF]{S.~Sjue}
\author[Mines]{D.~Smalley}
\author[TRIUMF]{C.~Sumithrarachchi}
\author[Guelph]{C.~E.~Svensson}
\author[TRIUMF,iThemba]{S.~Triambak}
\author[Liverpool,TRIUMF]{C.~Unsworth}
\author[York]{R.~Wadsworth}

\address[Surrey]{Department of Physics, University of Surrey, Guildford, Surrey GU2 7XH, UK.}
\address[LPC]{LPC, ENSICAEN, CNRS/IN2P3, UNICAEN, Normandie Universit\'e, 14050 Caen Cedex, France}
\address[York]{Department of Physics, University of York, York, YO10 5DD, UK}
\address[TRIUMF]{TRIUMF, 4004 Wesbrook Mall, Vancouver, BC, V6T 2A3, Canada}
\address[Liverpool]{Department of Physics, University of Liverpool, Liverpool, L69 3BX, UK}
\address[StMary]{Department of Physics, St Mary's University, Halifax, NS, B3H 3C3, Canada}
\address[LSU]{Department of Physics, Louisiana State University, Baton Rouge, LA 70803, USA}
\address[Toronto]{Department of Physics, University of Toronto, Toronto, Ontario, M5S 1A7, Canada}
\address[Mines]{Department of Physics, Colorado School of Mines, Golden, CO 80401, USA}
\address[Guelph]{Department of Physics, University of Guelph, Guelph, ON, N1G 2W1, Canada}
\address[UWC]{Department of Physics, University of the Western Cape, P/B X 17 Bellville, 7535, South Africa.}
\address[iThemba]{The iThemba Laboratory for Accelerator Based Sciences, Somerset West 7129, South Africa.}

\fntext[corresponding]{Corresponding author at Department of Physics, University of Surrey, Guildford, Surrey GU2 7XH, UK.}
%at Department of Physics, University of Surrey, Guildford, GU2 7XH, UK.}
\fntext[mines]{Now at Department of Physics, Colorado School of Mines, Golden, CO 80401, USA.}

\begin{abstract}

The levels in $^{26}$Na with single particle character have been observed for the first time using the d($^{25}$Na,p$\gamma$)
reaction at 5 MeV/nucleon.
The measured excitation energies and the deduced spectroscopic factors
are in good overall agreement with (0+1)$\hbar\omega$ shell model calculations performed in a complete $spsdfp$ basis and incorporating
a reduction in the $N=20$ gap.  Notably,
the $1p_{3/2}$ neutron configuration was found to play an enhanced role in the structure of the low-lying negative parity states in $^{26}$Na, compared to the isotone $^{28}$Al. Thus, the lowering of
the $1p_{3/2}$ orbital relative to the $0f_{7/2}$ occuring in the neighbouring
$Z$=10 and 12 nuclei -- $^{25,27}$Ne and $^{27,29}$Mg -- is seen also to occur at $Z$=11 and further strengthens the constraints on the modelling of the transition into the island of inversion.

\end{abstract}


\end{frontmatter}

% \linenumbers


The breakdown of the shell model magic number $N=20$ for neutrons, in the case of neutron-rich isotopes near $^{31}$Na and $^{32}$Mg \cite{Thibault,Detraz}, has led to the concept of an ``island of inversion'', where neutrons preferentially occupy orbitals above the normal $N=20$ gap, leaving vacancies below it \cite{WBB,Orr}. The mechanism is now understood in terms of the shell model to involve specific valence nucleon interactions \cite{WBB,PR} as well as a monopole shift in the effective single particle energies \cite{PR2}. In a wider picture, the monopole migration of levels is understood as arising from the tensor and three-body components of the nucleon-nucleon force between valence nucleons in partially filled orbitals \cite{Otsuka1,Otsuka2}.

\begin{figure}[c]
\includegraphics[width=1.0\textwidth]{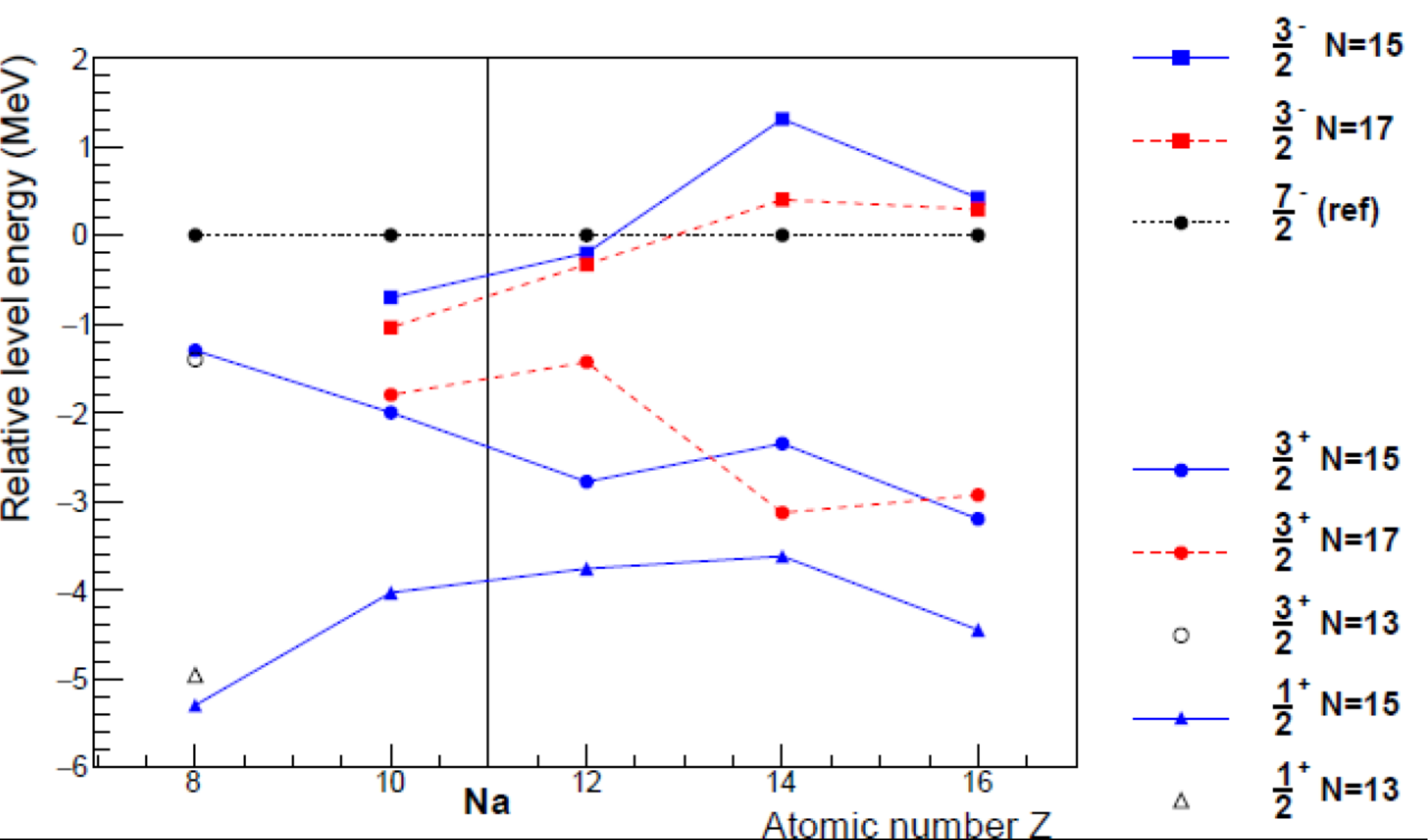}
\caption{(Colour online) Energies of lowest levels in isotones $N=15$ (dashed line) and $N=17$ (full line). Energies for $N=13$ are included for oxygen. The points for Z=10-16 are derived from refs. \protect\cite{Baumann, Obertelli, Terry, Catford_Ne25, Brown} and compilations, whilst those for oxygen are from refs. \protect\cite{Fernandez, Elekes}. The energy reference is chosen to be the $7/2^-$ level (see text) and the $Z$ position of $^{26}$Na is indicated. }
\label{systematics}       % Give a unique label
\end{figure}

The level schemes of nuclei with odd $N$, such as $N=15$ and $N=17$ (Figure \ref{systematics}), provide important insight into the evolution of the single particle structure approaching the island of inversion. Between $Z=8$ and $14$, the proton $d_{5/2}$ orbital is filled, and this affects the energies of the neutron orbitals. The reference energy is the $7/2^-$ level, as it is expected that the energy of the $f_{7/2}$ neutron orbital (a $j_>$ orbital, i.e. $j=\ell + 1/2$) will depend relatively weakly on the filling of the proton $d_{5/2}$ orbital (also $j_>$ \cite{Otsuka1}). In the case of $^{21,23}$O the $J^\pi = 7/2^-$ assignments are tentative \cite{Fernandez, Elekes} but the small width observed in $^{23}$O \cite{Elekes} supports the $7/2^-$ assignment. The systematics and also the shell model \cite{Elekes} suggest a near degeneracy between the $3/2^+$ and $3/2^-$ states in oxygen. As may be seen, $^{26}$Na falls in a key region where the shell closures at $N=20$ and $16$ are weakening and strengthening respectively.  As such, compared to the isotone $^{28}$Al, the levels in $^{26}$Na corresponding to a neutron in the $0d_{3/2}$ orbital
would be expected to lie higher in energy, whereas states with a neutron occupying the $1p_{3/2}$ orbital may be degenerate with the $0f_{7/2}$
levels or even lie below them\footnote{A multiplet of levels for each neutron configuration is expected, according to the coupling of the neutron with the unpaired proton in the $0d_{5/2}$ orbital.}.  In order to explore this behaviour the present study has been undertaken to locate the corresponding levels using neutron transfer onto $^{25}$Na, produced as a secondary beam. As will be shown, the detection of coincident gamma-rays has dramatically enhanced the excitation energy resolution and provided vital complementary information on the spins and parities of the populated levels.

Previous studies have identified the ground state of $^{26}$Na as having $J^\pi = 3^+$ \cite{Klotz,Alberger} and there are three $1^+$ states fed by the beta-decay of $^{26}$Ne \cite{Weissman,FSU}. Apart from the low-lying quartet of states ($3^+, 1^+, 2^+, 2^+$) below 407 keV \cite{Nobby1}, which have been very recently probed via Coulomb excitation \cite{coulex}, the many observed states \cite{Nobby1,Nobby,Errol,FSU} have a structure that is almost completely unknown. The (d,p) reaction employed here will selectively populate the levels with a predominantly single-particle structure, which are those most useful for testing shell model predictions.

The experiment was undertaken at the ISAC-II facility at TRIUMF. A pure beam of 5.0 MeV/nucleon $^{25}$Na ions of intensity $3 \times 10^7$ pps was employed to bombard a self-supporting (CD$_2$)$_n$ deuterated polythene foil of thickness 0.5 mg/cm$^{2}$. The target was mounted at the centre of the SHARC silicon strip detector array \cite{SHARC} which was surrounded by 8 segmented germanium clover gamma-ray detectors of the TIGRESS array \cite{TIGRESS}. Four clovers were centred at $90^\circ$ in the laboratory and four at $135^\circ$, giving a total absolute photopeak efficiency of some 3\% at 1.33 MeV. In order to identify events in SHARC arising from fusion-evaporation reactions initiated on the carbon in the target (a source of significant background), an Al foil followed by a thin plastic scintillator detector (the TRIFOIL \cite{Benoit-thesis}) were installed 400 mm downstream of the target. The Al thickness of 30~$\mu$m was chosen to stop fusion-evaporation residues whilst transmitting the beam and the products of direct reactions \cite{Trifoil}.  Further, the scintillator foil was thin enough (10~$\mu$m) to transmit the radioactive beam.

Events were recorded whenever a particle was registered in SHARC, including a logic signal to indicate whether the TRIFOIL had fired during the same beam bunch. The TRIFOIL allowed the events corresponding to the production of $^{26}$Na via the (d,p) reaction to be highlighted in the analysis in order to optimise the software gating. In the final determination of the absolute differential cross sections of the (d,p) reaction, the TRIFOIL requirement was not imposed, however, owing to its efficiency being dependent on the position on the foil, and hence on the angle of the recoil proton \cite{Wilson_thesis}. The coincident gamma-rays were recorded in TIGRESS and their energies were corrected for Doppler shift ($v/c \approx 0.1$) \cite{Wilson_thesis}.

\begin{figure}[c]
\includegraphics[width=.7\textwidth]{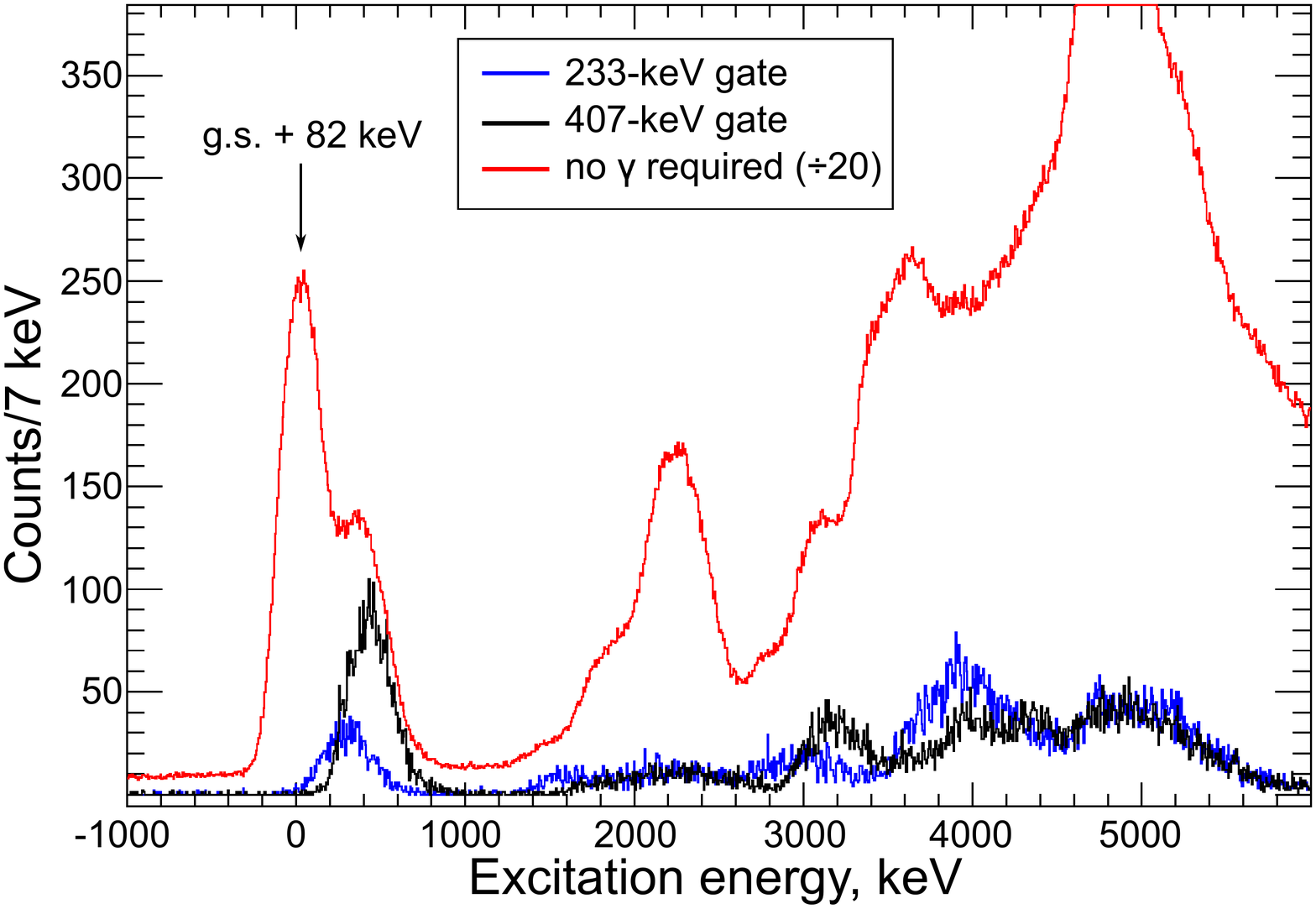}
\caption{(Colour online) Excitation energy spectrum for states in $^{26}$Na, reconstructed from the measured energy and angle of the protons from the d($^{25}$Na,p) reaction. The spectrum for all events is shown (red), plus the results for coincidences with the known gamma-rays of 233 keV (blue) and 407 keV (black histogram) in $^{26}$Na. The data are for proton laboratory angles backward of 90$^\circ$. A TRIFOIL requirement has been imposed in all cases (see text).}
\label{excitation}       % Give a unique label
\end{figure}

The excitation energy spectrum for states populated in $^{26}$Na, reconstructed from the observed energy and angle of the protons, is shown in Figure \ref{excitation} where the resolution is 350 keV (FWHM). This figure illustrates how closely spaced states could be distinguished using the gamma-ray data.
The FWHM resolution in gamma-ray energy at 1.8 MeV, after Doppler correction, was 18 keV at 135$^\circ$ and 23 keV at 90$^\circ$. This resolution, rather than that of the proton-derived excitation energy spectrum, defined the precision with which individual states could be selected. On the other hand, the proton-derived excitation energy was critical in determining the energy at which the $^{26}$Na was populated in the (d,p) reaction (i.e. prior to gamma-decay of the states). In addition, gating on this excitation energy was employed in the determination of the gamma-ray branching ratios \cite{Wilson_thesis}.

Elastically scattered deuterons from the target were prominent in the spectrum of particle energy versus angle and the corresponding differential cross section was extracted \cite{Wilson_thesis}. Optical model parameters for $d+^{26}$Mg at an almost identical centre of mass energy  \cite{Meurdersdd} were employed to construct a theoretical angular distribution and this was found to reproduced well the form of the measured distribution (including the observed minimum). The best-fit normalisation thus gave the product of the deuteron target thickness and the total integrated beam flux. The uncertainty associated with this fitting was estimated to be 3\%. An estimate of systematic uncertainties in this analysis was obtained by repeating the procedure with parameters taken from $d+^{28}$Si scattering at the same energy \cite{Yule}. The variation was less than 2\%.

\begin{figure}[c]
\includegraphics[width=.9\textwidth]{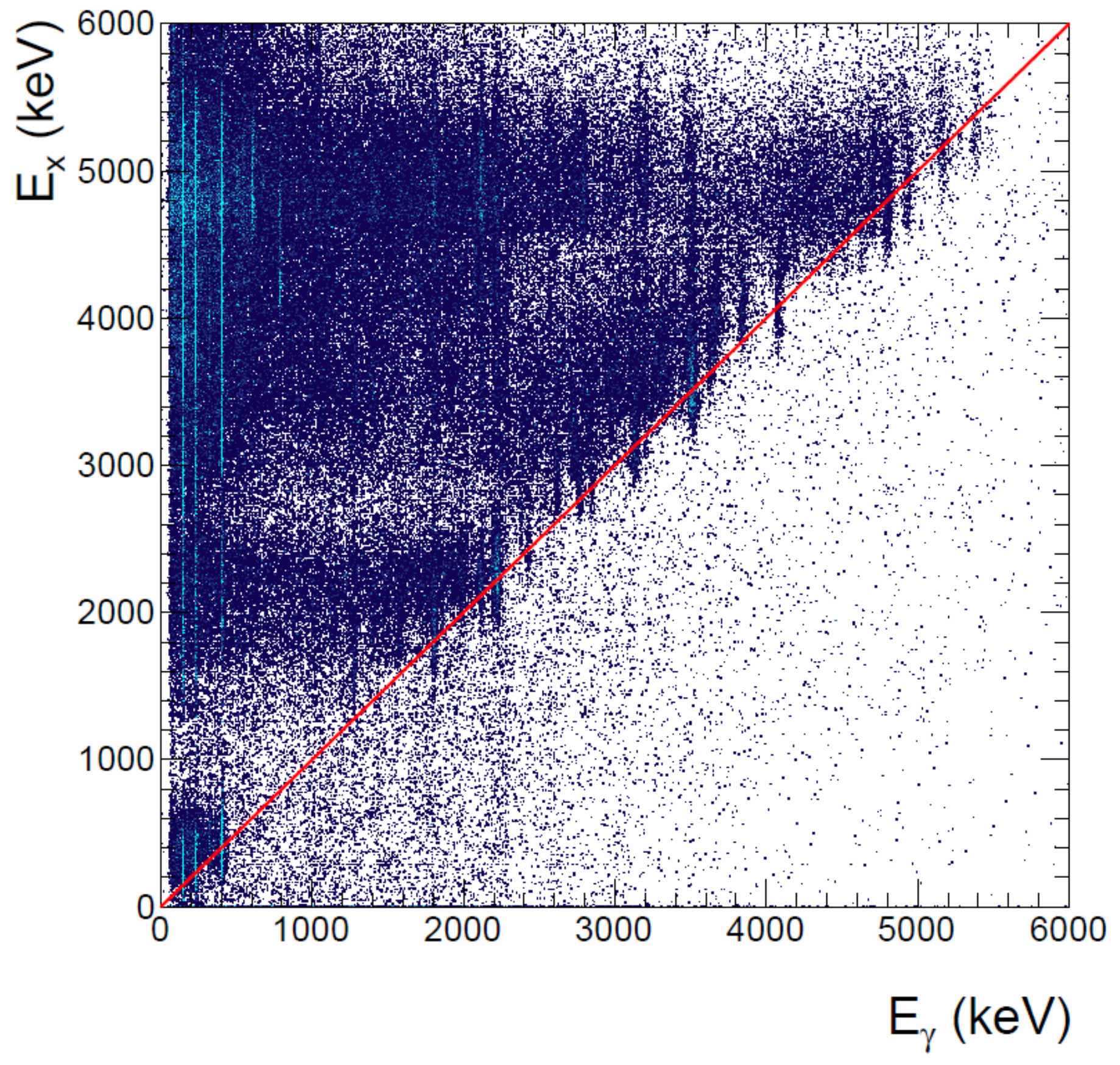}
\caption{(Colour online) Plot of the excitation energy $E_{\rm x}$ of $^{26}$Na states populated via the d($^{25}$Na,p) reaction, as deduced from the proton energy and angle, versus the energy of any coincident gamma-ray (after Doppler correction), $E_{\gamma}$. Gamma-decays directly to the ground state lie along the diagonal $E_{\rm x}$ = $E_{\gamma}$, whilst cascade decays result in events lying above the diagonal. Data are for laboratory proton angles backward of 90$^\circ$. A TRIFOIL requirement has been imposed (see text). }
\label{money}       % Give a unique label
\end{figure}

Figure \ref{money} shows the Doppler corrected gamma-ray energy ($E_{\gamma}$) plotted against the excitation energy in $^{26}$Na ($E_{\rm x}$) as derived from the proton energy and angle. The superior resolution of the gamma-ray energy is clearly apparent, and the states that overlap in $E_{\rm x}$ can be distinguished using the gamma-ray energy. A number of levels with ground-state gamma-ray branches ($E_{\gamma}=E_{\rm x}$) can be seen, including many of the more strongly populated states. The neutron separation energy is $S_n = 5.57$ MeV \cite{ame2012}. The ground state branch generally has a lower and better defined underlying background than any other peak in Figure \ref{money} and hence for the most part only these peaks were chosen for initial analysis\footnote{A full analysis including cascade decays and the more weakly populated states will follow.} (Table 1). With suitable background subtraction \cite{Wilson_thesis}, the yield of protons could be deduced for individual states in $^{26}$Na as a function of laboratory angle. These distributions were converted to absolute differential cross sections by taking into account the geometry of SHARC, the elastic scattering normalisation, the measured gamma-ray branching ratios and the gamma-ray detection efficiency (corrected for Doppler and relativistic angular abberation effects). The overall systematic error associated with these effects was typically 5-6\%.

\begin{figure}[c]
\includegraphics[width=\textwidth]{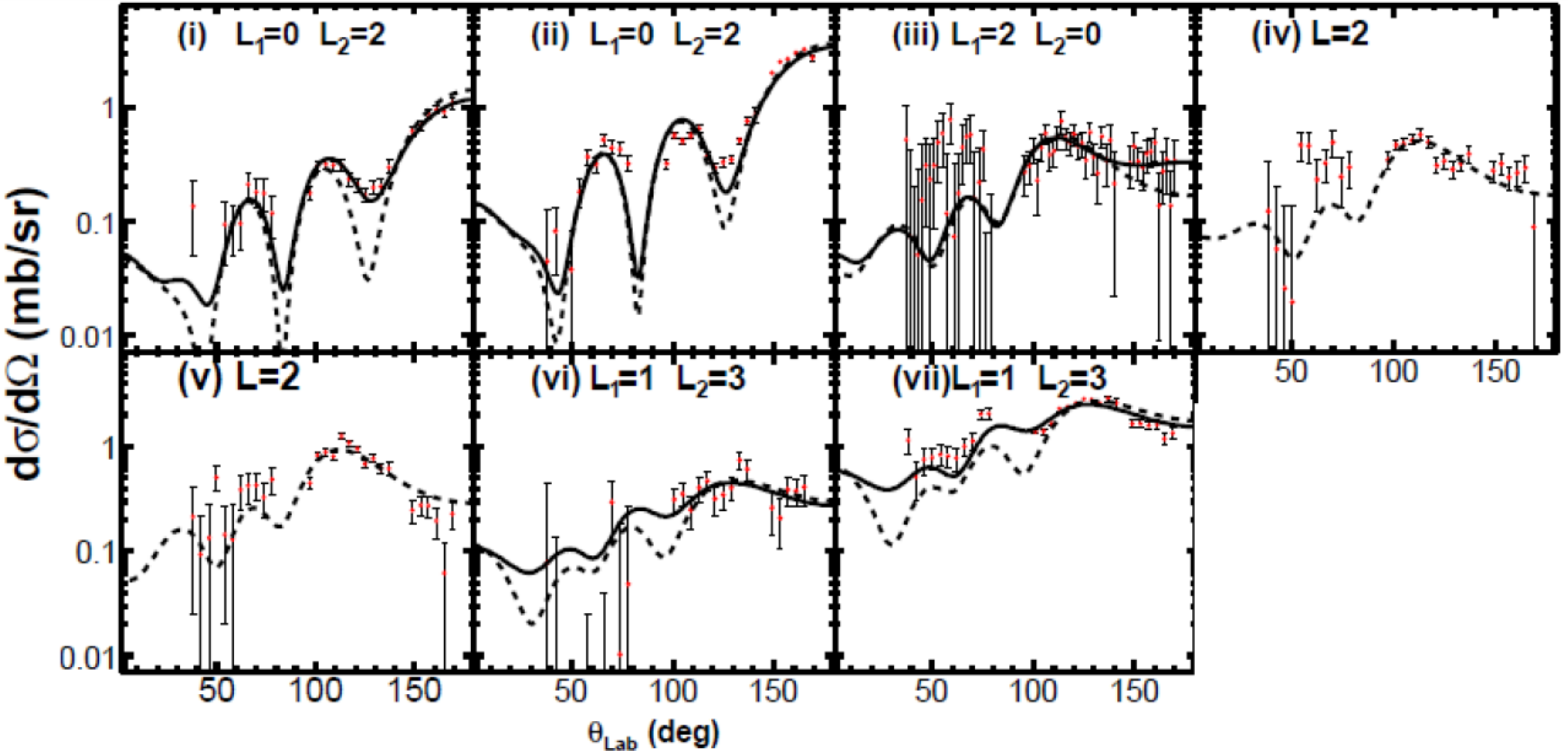}
\caption{Differential cross sections for the reaction d($^{25}$Na,p) at 5.0 MeV/nucleon. The results of ADWA reaction calculations are also shown, normalised to the data, for the angular momentum transfers
indicated and listed in Table 1 (dashed line = fit with single angular momentum, labelled as L or as L$_1$ where another fit is also shown, full line = sum of two contributions L$_1$ and L$_2$).}
\label{differential}       % Give a unique label
\end{figure}

Differential cross sections for states populated in $^{26}$Na are shown in Figure \ref{differential}, plotted in terms of the laboratory scattering angle. These are compared with reaction calculations performed using the code TWOFNR \cite{TWOFNR} employing the Adiabatic Distorted Wave Approximation (ADWA) of Johnson and Soper \cite{Johnson} with standard input parameters \cite{Lee} including the Chapel-Hill (CH89) nucleon--nucleus optical potential \cite{CH89}. This formalism benefits {\it inter alia} from having no requirement for a deuteron-nucleus optical potential. The magnetic substate populations from TWOFNR were used to check that the gamma-ray coincidence requirement did not alter the shape of the differential cross sections by more than a few percent across the full range of proton angles \cite{Wilson_thesis}. Spectroscopic factors, $S$, were extracted by normalising the calculated differential cross sections to the data using the full range of angles shown in Figure \ref{differential}. The transferred angular momentum, L, that best describes the shape of the measured angular distribution for each state is listed in Table 1.

States for which more than one value is possible for the transferred angular momentum have been analysed by fitting a linear combination of the calculated cross sections using the two possible values L$_1$ and L$_{2}$ (where L$_1$=L). It is important to note that the contribution from the larger of L$_1$ and L$_2$ is suppressed (for a similar spectroscopic factor) due to poorer kinematic matching. As a result, the spectroscopic factor deduced for the higher L is seen generally to exhibit a larger statistical uncertainty. As may be seen in Table 1, the addition of L$_2$ to the fit does not significantly change the spectroscopic factor deduced using just L$_1$, i.e. $S_1 \approx S$. The overall uncertainties assigned to the spectroscopic factors are some 20\% (dominated overwhelmingly by the reaction calculations) \cite{Lee}. The inferred spin and parity assignments are discussed below.

%% Oh my god its sideways ... I'm getting Dizzy, I may be sick ...

\begin{sidewaystable}
\label{SFs}
%\begin{table}
\caption{Excitation energies (E$_{\rm x}$) and proposed $J^\pi$ for $^{26}$Na levels in the present work, with angular momentum transfers and deduced spectroscopic factors ($\pm 20$ \%)$^{\S}$. The L-transfer dominating the yield is denoted by L, or L$_1$ where two L-values are possible; the second value is then denoted by L$_2$. Values of $S$ and $S_{1,2}$ were obtained by fitting cross sections using a single L and a sum of two L contributions, respectively, assuming orbitals $(n l j)$. Shell model spectroscopic factors $S^{SM}$ and $S^{SM}_{1,2}$ are shown for each fitted orbital $(n l j)$ and, for completeness, its spin-orbit partner also. Numbering of states as in Figure \ref{differential}.}

\setlength{\tabcolsep}{3.pt}
\begin{tabular}{cccccccccccccccccccc} \hline
&\hspace{0.0mm}	&\hspace{0.0mm}	&\hspace{0.0mm}	&\hspace{0.0mm}	& \multicolumn{6}{c}{single L analysis} & \multicolumn{9}{c}{two L analysis (where applicable)} \\
\cline{7-10} \cline{12-20}
No. & E$_{x}$ $^{a)}$  &  E$_{x}^{SM~b)}$ &J$^{\pi~c)}$ & J$^\pi_{SM}$& & L & $nlj$ & S & S$^{SM}$ & & L$_1$ & $n_1l_1j_1$ & S$_1$ & S$_1^{SM}$  & & L$_2$ & $n_2l_2j_2$ & S$_2$ & S$_{2}^{SM}$ \\ \hline\hline
~ &  0  & 0       &  3$^+$ & 3$^+_1$ & &  * & 1s$_{1/2}$ & & 0.61 & & * & 1s$_{1/2}$ & & 0.61 &  & * & 0d$_{3/2}$ & & 0.01 \\
  &     &         &        &         & &    &            & &      & &   &            & &      &  &   & 0d$_{5/2}$ & & 0.01 \\
~ & 0.082$^d)$  & 0.077      & 1$^+$  & 1$^+_1$ & &  * & 0d$_{3/2}$ & & 0.29 & &   &            & &      &  &   &            & &      \\
  &     &         &        &         & &    & 0d$_{5/2}$ & & 0.11 & &   &            & &      &  &   &            & &      \\
(i) & 0.232 & 0.149 & 2$^+$ & 2$^+_1$ & & 0 & 1s$_{1/2}$ & 0.13 & 0.15 & & 0 & 1s$_{1/2}$ & 0.10 & 0.15  & &  2 & 0d$_{3/2}$ & 0.19\dag & 0.10 \\
    &       &       &       &         & &   &            &      &      & &   &            &      &       & &    & 0d$_{5/2}$ &       & 0.09 \\
(ii) & 0.405 & 0.416 & 2$^+$ & 2$^+_2$ & & 0 & 1s$_{1/2}$ & 0.33 & 0.27 & & 0 & 1s$_{1/2}$ & 0.30 & 0.27 & & 2 & 0d$_{5/2}$ & 0.13\dag & 0.03\\
     &       &       &       &         & &   &            &      &      & &   &            &      &      & &   & 0d$_{3/2}$ &       & 0.03 \\
~  & 1.507 & 1.409 & 1$^+$ & 1$^+_2$ & & * & 0d$_{3/2}$ &          & 0.09 & &   &            &        &      & &   &            &          &     \\
   &       &       &       &         & &   & 0d$_{5/2}$ &          & 0.10 & &   &            &        &      & &   &            &          &     \\
(iii) & 1.805 & 1.676 & (3$^+$) & 3$^+_2$ & & 2 & 0d$_{3/2}$ & 0.37 & 0.33 & & 2 & 0d$_{3/2}$ & 0.33\dag & 0.33 & & 0 & 1s$_{1/2}$ & 0.01\ddag & 0.00\\
      &       &       &       &         & &   & 0d$_{5/2}$ &      & 0.02 & & 2 & 0d$_{5/2}$ &       & 0.02 & &   &            &        &      \\
(iv) & 2.116 & 2.241 & 5$^+$ & 5$^+_1$ & & 2 & 0d$_{5/2}$ & 0.16 & 0.08 & &  &  &  &  & &  &  &  &  \\
(v) & 2.225 & 2.048 & (4$^+$) & 4$^+_2$ & & 2 & 0d$_{3/2}$ & 0.43 & 0.51 & &  &  &  &  & &  &  &  &  \\
    &       &       &       &         & &   & 0d$_{5/2}$ &      & 0.01 & &   &   &   &   & &   &   &   &  ~ \\
~  & 2.843 & 2.936 & (2$^-$) & 2$^-_1$ & & * & 0f$_{7/2}$ &          & 0.20 & & * & 0f$_{7/2}$ &         & 0.20 & & * & 1p$_{3/2}$ &         & 0.05\\
   &       &       &       &         & &   & 0f$_{5/2}$ &          & 0.00 & &   & 0f$_{5/2}$ &         & 0.00 & &   & 1p$_{1/2}$ &         & 0.04 \\
(vi) & 3.135 & 3.228 & 3$^-$ & 3$^-_1$ & & 1 & 1p$_{3/2}$ & 0.07\dag & 0.15 & & 1 & 1p$_{3/2}$ & 0.06\dag & 0.15 & & 3 & 0f$_{7/2}$ & 0.10\ddag & 0.13\\
     &       &       &       &         & &   & 1p$_{1/2}$ &       & 0.02 & &   & 1p$_{1/2}$ &       & 0.02 & &   & 0f$_{5/2}$ &        & 0.00 \\
(vii) & 3.511 & 3.513 & 4$^-$ & 4$^-_1$ & & 1 & 1p$_{3/2}$ & 0.30 & 0.44 & & 1 & 1p$_{3/2}$ & 0.25 & 0.44 & & 3 & 0f$_{7/2}$ & 0.51\dag & 0.00\\
      &       &       &       &         & &   &            &      &      & &   &            &      &      & &   & 0f$_{5/2}$ &       & 0.00\\
~  & 4.305 & 4.401 & (5$^-$) & 5$^-_1$ & & * & 0f$_{7/2}$ &          & 0.46 & &   &            &        &      & &   &            &          &     \\
   &       &       &         &         & &   & 0f$_{5/2}$ &          & 0.00 & &   &            &        &      & &   &            &          &     \\
~  & 4.917 & 4.881 & (6$^-$) & 6$^-_1$ & & * & 0f$_{7/2}$ &          & 0.61 & &   &            &        &      & &   &            &          &     \\
~  & 5.009 &       & (3$^-$,4$^-$) &   & & * &   &          &   & &   &            &        &      & &   &            &          &     \\ \hline
\end{tabular}

\begin{footnotesize}
\begin{tabular}{rl}
\S     & dominated by reaction theory contribution; statistical errors typically several percent except where noted as 10\% (\dag) or 35\% (\ddag) \\
\verb+*+      & extraction of these differential cross sections beyond scope of present analysis (see text) \\
$^{a)}$ & present work, excitation energy in MeV ($\pm 1$ keV) deduced from Doppler-corrected gamma-ray energies\\
$^{b)}$ & excitation energy in MeV, calculated using the shell model (SM) interactions USD-A ($\pi = +$) \cite{USDAB} and WBP-M ($\pi = -$) \cite{Brown} (see text)\\
$^{c)}$ & inferred in present work (see text) except g.s. \cite{Klotz} and first four excited states \cite{Nobby1,Weissman,FSU}\\
$^{d)}$ & excitation energy of this isomeric state deduced from gamma-decays feeding both this level and the g.s. (see Figure \ref{levels})\\
\end{tabular}
\end{footnotesize}
%%\end{table}
\end{sidewaystable}

\begin{figure}[c]
\includegraphics[width=.7\textwidth]{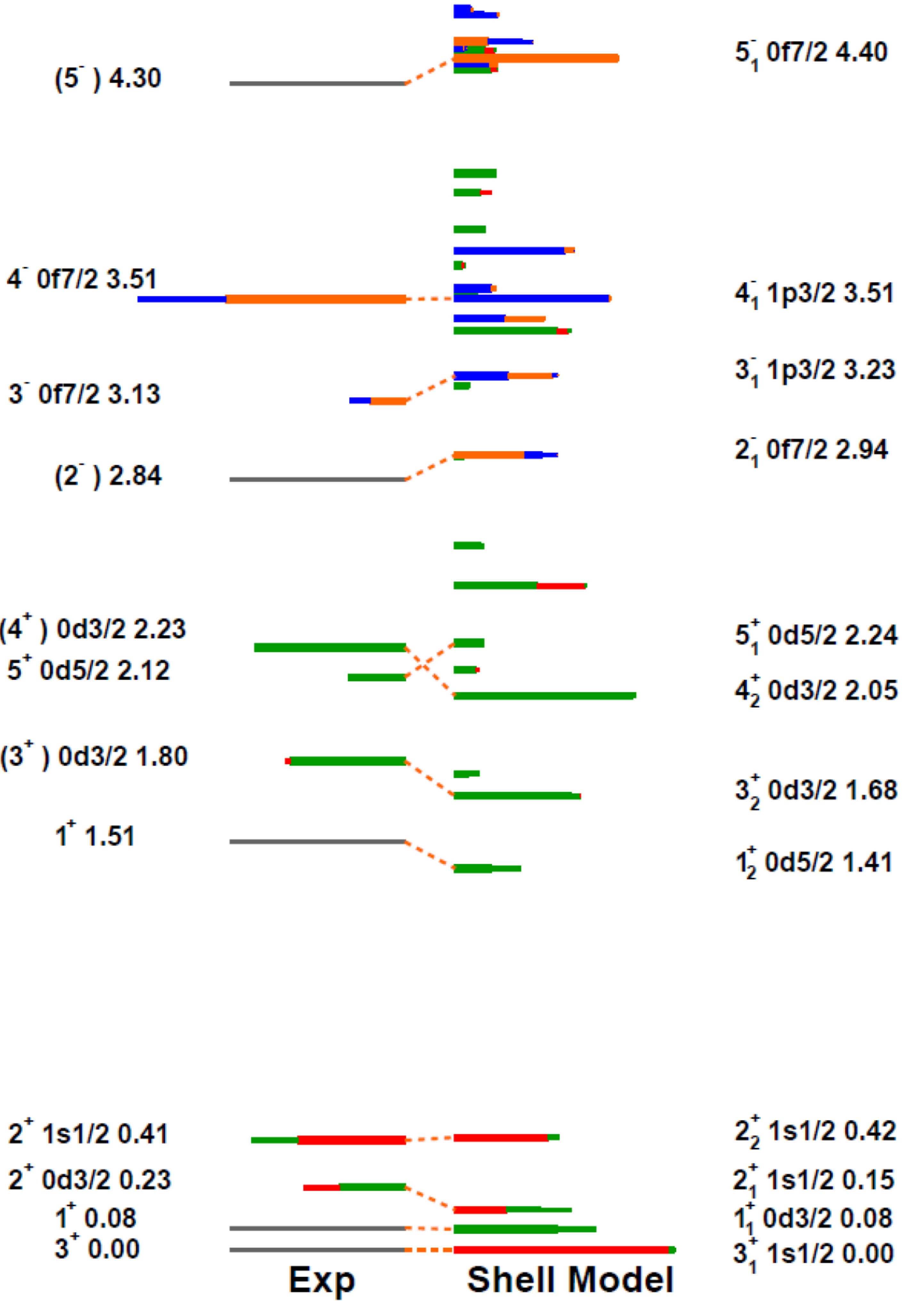}
\caption{(Colour online) Level scheme of $^{26}$Na as deduced from the present work compared to the results of shell model calculations
employing the USD-A and WBP-M interactions (see text). The lengths of the coloured lines correspond to the spectroscopic factors as follows: red $s$-wave, blue $p$-wave, green $d$-wave, orange $f$-wave.}
\label{shell-model}       % Give a unique label
\end{figure}

\begin{figure}[c]
\includegraphics[width=.7\textwidth,angle=90]{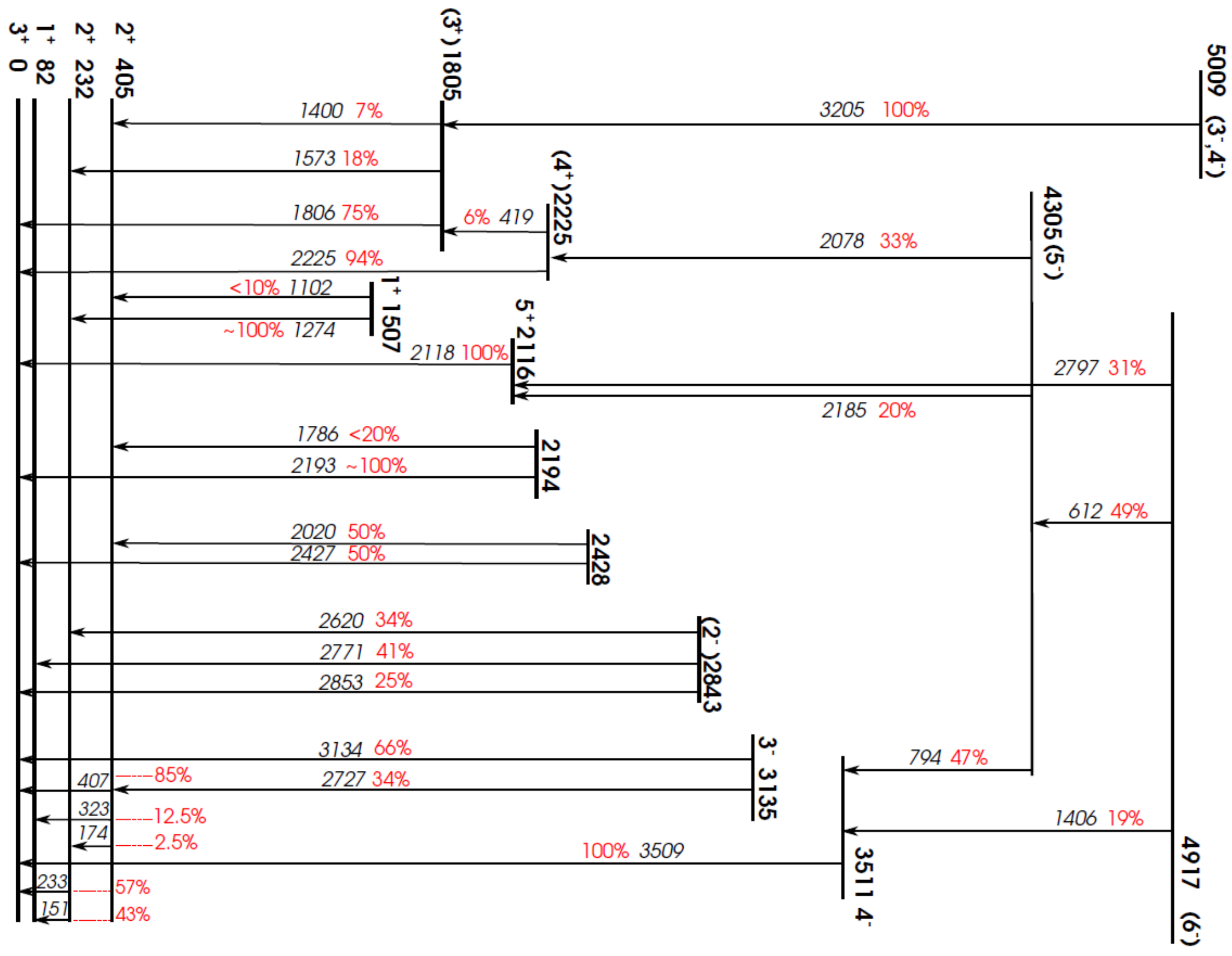}
\caption{(Colour online) Decay scheme of $^{26}$Na as deduced from the present work, including the gamma-ray branching ratios.}
\label{levels}       % Give a unique label
\end{figure}

The $^{25}$Na projectile has a structure, in the simplest shell model picture, of three $0d_{5/2}$ protons coupled to $5/2^+$ and a closed $0d_{5/2}$ neutron orbital. The positive parity orbitals $1s_{1/2}$ and $0d_{3/2}$, as well as the negative parity orbitals $0f_{7/2}$, $1p_{3/2}$, $1p_{1/2} \ldots$ are thus available for neutron transfer. As shown in Figure \ref{excitation}, the known low-lying positive parity states below 450 keV \cite{Klotz,Nobby1} are populated.  As discussed below, the data also show clearly that there is population of negative parity states, as would be expected from the previous (d,p) studies of $^{25,27}$Ne \cite{Catford_Ne25,Brown}, $^{27}$Mg \cite{Meurders} and $^{28}$Al \cite{Freeman70,Meurders,Freeman}. In order to interpret the results, shell model calculations have been performed including all $0\hbar \omega$ $sd$-shell configurations for positive parity states and all $1\hbar \omega$ excitations in an $spsdpf$ basis for negative parity states. The program OXBASH \cite{OXBASH} was used with the USD-A \cite{USDAB} and WBP-M \cite{Brown} interactions respectively. The WBP-M interaction, which describes the neighbouring nuclei $^{25,27}$Ne and $^{29}$Mg in a consistent manner \cite{Brown}, is a modification of the WBP interaction \cite{WBP} that shifts the energies of the $fp$-shell orbitals down by 0.7 MeV. It was previously noted \cite{Brown} that for the neon and magnesium isotopes the strongly populated $3/2^-$ and $7/2^-$ states ($S \approx 0.4$-$0.6$) appear to track fairly closely the shift in the energy of the $fp$-shell orbitals. In fact, earlier work by Bender and coworkers also applied a similar interaction called WBP-a that improved the agreement with the shell model for states in nearby (and only slightly higher $Z$) isotopes of Al \cite{Bender1} and P \cite{Bender2,Bender3}.
Excitation energies for the WBP-M and USD-A calculations were computed relative to the (positive parity) ground state calculated with the same interaction. The association of experimentally observed states with shell model states, indicated in Table 1 and Figure \ref{shell-model}, was based on the observed L-transfer, the excitation energy and the gamma-decay selectivity (Figure \ref{levels}) as discussed below.

Above $\sim$2.5 MeV excitation, no states of positive parity are expected to be strongly populated, according to the shell model calculations. The states seen here at higher energies are therefore likely to have negative parity. Of these, the most easily identified is the strongly populated state at 3.511 MeV which is observed to have a yield dominated by L$=1$ neutron transfer at large laboratory angles (Figure \ref{differential}). The only observed gamma-decay branch for this state is to the $3^+$ ground state, in line with the favoured decay pattern for the lowest $4^-$ state in the isotone $^{28}$Al \cite{Freeman70}. This state is thus assigned to be the $4^-_1$ state in $^{26}$Na.

There are two strongly populated higher lying levels that gamma-decay via the 3.511 MeV level (Figure \ref{levels}). In terms of predicted levels in this energy range, the obvious candidates are the lowest $5^-$ and $6^-$ states (Table 1). These could reasonably be expected to gamma-decay via the $4^-$ state, according to the pattern observed \cite{Freeman} in $^{28}$Al.

The decays of both the proposed $5^-$ and $6^-$ states proceed in part (20\% and 31\% respectively \cite{Wilson_thesis}) via the state at 2.116 MeV. Given that this level is populated via an L$=2$ transfer (Figure \ref{differential}) and that it is fed by these higher lying negative parity states, it has a likely $J^\pi$ assignment of greater than $4$, and a comparison of possible spins with shell model calculations indicates an assignment of $5^+$ (supported by both the energy and the weak spectroscopic factor). Indeed, there is an analogous state at 2.581 MeV in $^{28}$Al, perhaps weakly populated in the decay of the $6^-$ state \cite{Freeman}, that has a $5^+$ assignment \cite{Maher,Endt}.

The two reasonably strong states seen just below the $4^-$ are most naturally associated with the $2^-$ and $3^-$ states predicted in the shell model (see Table 1). In particular, the identification of the $3^-$ at 3.135 MeV is confirmed by its gamma-decay branch to the lowest lying $1^+$ state at 1.507 MeV as
occurs in $^{28}$Al (Table 28.6 of ref. \cite{Endt}). This is further supported by the mixed L$_1=1$ plus L$_2=3$ differential cross section (Figure \ref{differential}).

The spectroscopic factors extracted from the proton differential cross sections are compared to the shell model predictions in Table 1 and Figure \ref{shell-model}. The $1s_{1/2}$ strength that leads to $2^+$ states is concentrated in the two known states below 0.5 MeV and sums to approximately 0.45 in both theory and experiment. The tentative identifications of $3_2^+$ and $4_2^+$ states are based on a comparison with the shell model calculations as follows.
The observed $0d_{3/2}$ strength leading to $J^\pi =3^+$ is concentrated in one state near 1.8 MeV with a spectroscopic strength in very good agreement with the predictions. The $0d_{3/2}$ strength leading to the $4^+$ states has a comparable magnitude in theory and experiment and is located close to 2 MeV, so that the experiment is compatible with the strength being concentrated in the second $4^+$ state. We note that the USD-A interaction \cite{USDAB} predicts the excitation energies for positive parity states to within $\sim$150 keV, in line with typical shell model accuracy.
When we instead used the WBP-M interaction to calculate the energies of states with a strong $0d_{3/2}$ neutron character (namely the $3^+$ and $2^+$ states), then the predictions lay about 0.4 MeV lower than experiment. This is because the WBP-M incorporates the USD interaction \cite{USD} to compute the $0\hbar \omega$ positive parity levels and this is known to underestimate $0d_{3/2}$ neutron energies \cite{Catford_Ne25}.

The spectroscopic factors deduced for the $4^-$ state indicate a comparable strength to that predicted for the $1p_{3/2}$ transfer, albeit somewhat weaker. However, the shell model fails completely to reproduce the substantial $0f_{7/2}$ strength. The excitation energy is given accurately by the WBP-M calculation, as are those of the other negative parity states, within 100 keV ($J^\pi = 2^-, 3^-, 5^-, 6^-$). These latter states are all predicted to have a structure that overlaps substantially with that of a neutron in the $0f_{7/2}$ orbital coupled to the $0d_{5/2}$ proton of the $^{25}$Na ground state. It is worthwhile noting that the $3^-$ state is predicted also to have a similar spectroscopic strength for the coupling with a $1p_{3/2}$ neutron. Because of a better kinematic matching, the L$=1$ transfer dominates the yield; a similar situation is observed for the lowest $3^-$ state in $^{28}$Al as populated in (d,p) \cite{Carola,Chen,Maher}. In $^{28}$Al, the $0f_{7/2}$ measured spectroscopic factor actually exceeds that of the $1p_{3/2}$ orbital by a factor of three \cite{Carola} (Table 2). Another $3^-$ state in $^{26}$Na is predicted by the WBP-M calculations to have approximately equal mixing and to lie at 4.462 MeV, whilst a predominantly $1p_{3/2}$ neutron state occurs at 4.774 MeV. These states may be populated but have not been identified in the present analysis.

\begin{table}
\label{negpar}
\caption{Mixing between 1p$_{3/2}$ and 0f$_{7/2}$ configurations in $^{26}$Na, compared with the isotone $^{28}$Al. Values of (2J+1)S from \cite{Carola} are converted to S using the confirmed spins \cite{Endt}.
}

\begin{tabular}{cccccccc} \hline
& \multicolumn{3}{c}{$^{26}$Na, present work} &~& \multicolumn{3}{c}{$^{28}$Al, ref. \protect{\cite{Carola}}} \\
\cline{2-4} \cline{6-8}
J$^\pi$  & E$_{x}$\,$^{a)}$ &  S$_1$(1p$_{3/2}$)\,$^{b)}$ &S$_2$(0f$_{7/2}$)\,$^{b)}$ &&  E$_{x}$\,$^{a)}$  &  S$_1$(1p$_{3/2}$) &S$_2$(0f$_{7/2}$) \\ \hline\hline
$3^-$    & 3.135           &    0.06 {\it 0.01}     &    0.10 {\it 0.03}    &&  3.591            &   0.19             &   0.60  \\
$4^-$    & 3.511           &    0.25 {\it 0.01}     &    0.51 {\it 0.05}    &&  3.465            &   0.11             &   0.62  \\
$3^-$    &                 &                    &                  &&  4.691            &   0.31             &   0.08  \\
$2^-$    &                 &                    &                  &&  4.766            &   0.41             &   0.15  \\
$2^-$    &                 &                    &                  &&  4.905            &   0.24             &   0.06  \\
$3^-$    &                 &                    &                  &&  5.134            &        $^{c)}$     &         \\
\hline
\end{tabular}

\begin{footnotesize}
%\begin{tabular}{rl}
$^{a)}$\,excitation energy in MeV ~~
$^{b)}$\,statistical error in italics ~~
$^{c)}$\,beyond E$_x$ range of \cite{Carola}.\\
%\end{tabular}
\end{footnotesize}
\end{table}

In the case of the $4^-$ state in $^{26}$Na, the deduced $1p_{3/2}$ spectroscopic factor is twice as large as that reported for $^{28}$Al whereas the $0f_{7/2}$ spectroscopic factor is effectively unchanged (Table 2). In contrast, for the $3^-$ state, the spectroscopic factors are three times and six times lower, respectively, than in $^{28}$Al \cite{Carola,Endt}. A clearer picture emerges from Table 2 if the relative magnitudes of the $1p_{3/2}$ and $0f_{7/2}$ spectroscopic factors for each of the $^{26}$Na states are compared with the behaviour in $^{28}$Al. For both the $3^-$ and $4^-$ states in $^{26}$Na, the $1p_{3/2}$ spectroscopic factor is half the magnitude of that for $0f_{7/2}$. In contrast, the analogous states in $^{28}$Al exhibit $1p_{3/2}$ spectroscopic factors that are three to five times smaller. This demonstrates an enhanced role emerging for the $1p_{3/2}$ orbital in the structure of the low-lying negative parity states in $^{26}$Na as compared to $^{28}$Al. Indeed, in the case of $^{28}$Al the spectroscopic factors (Table 2) indicate that the states with predominantly $1p_{3/2}$ structure lie in the region of 4.8 MeV, significantly above the $0f_{7/2}$-dominated states which are closer to 4.0 MeV (Table 2 of ref. \cite{Carola}). It would be very interesting if the spectroscopic factors for the higher lying $2^-$ and $3^-$ states in $^{26}$Na could be measured and compared. As is evident in Figure \ref{shell-model}, a theory-based comparison between the $1p_{3/2}$ and $0f_{7/2}$ strengths in $^{26}$Na indicates that the levels with a $1p_{3/2}$ structure are on average around 1~MeV below those with $0f_{7/2}$ structure (in fact, 0.83~MeV when weighted by the spectroscopic factors), an inversion that is in accord with the systematics of Figure \ref{systematics}.

In conclusion, as noted in the introduction and illustrated in Figure \ref{systematics}, the ordering of levels in nuclei with A $\simeq$ 25--30 evolves dramatically as they become more neutron-rich, driven largely by the interaction between protons in the $0d_{5/2}$ orbital and the valence neutrons. Importantly in this context, the results presented here confirm that the evolution is also manifest in $^{26}$Na. In particular, the low-lying $3^-$ and $4^-$ states in $^{26}$Na are found to exhibit an enhanced influence of the $1p_{3/2}$ neutron orbital, compared to the isotone $^{28}$Al. In addition the WBP-M shell model calculations, in which the $fp$-shell orbitals are lowered to reduce the $N=20$ gap by 0.7 MeV, succeed in reproducing the energies of the negative parity states in $^{26}$Na as they do in the neighbouring neon and magnesium isotopes \cite{Catford_Ne25,Brown}. From a theoretical perspective, a less {\it ad hoc} description of the transition into the island of inversion represents an interesting and important challenge.

%In addition, it should be noted that
%the interaction between the valence protons and neutrons is probed further by the energy splitting of the multiplets of states
%formed by each proton-neutron coupling.

The assistance of our late colleague and friend R.M. Churchman in preparing the experiment is gratefully acknowledged. The efforts of the ISAC operations team in supplying the $^{25}$Na beam are appreciated. Illuminating discussions with B.A. Brown are also acknowledged. We acknowledge support from the Science and Technologies Facility Council (UK) (grants ST/J000051/1, STFC-EP/D060575/1) and the Natural Sciences and Engineering Research Council of Canada; TRIUMF
is funded via a contribution agreement with the National Research Council of Canada.



%% References



\end{document}